\documentclass[11pt]{article}
\usepackage{amsmath}
\usepackage{amssymb}
\usepackage{amsfonts}
\usepackage{amsthm}
\usepackage{graphicx}
\usepackage{subfigure}
\usepackage{enumerate}
\usepackage{verbatim}
\usepackage{authblk}
\usepackage{setspace}
\usepackage{enumerate}
\usepackage{comment}
\usepackage{bm}
\usepackage{tikz}
\definecolor{lightgray}{gray}{0.8}
\begin{document}
\title{On an Extremal Hypergraph Problem Related to Combinatorial Batch Codes}
\author {{\bf Niranjan Balachandran}\\
Department of Mathematics,\\
Indian Institute of Technology Bombay,\\ Mumbai.\\
\and
{\bf Srimanta Bhattacharya}\\
Centre of Excellence in Cryptology,\\
Indian Statistical Institute,\\ Kolkata.
}
\date{}
\maketitle
\newtheorem{theorem}{\bf Theorem}[section]
\newtheorem{proposition}[theorem]{\bf Proposition}
\newtheorem{lemma}[theorem]{\bf Lemma}
\newtheorem{corollary}[theorem]{\bf Corollary}
\newtheorem{definition}{\bf Definition}[section]
\newtheorem*{claim}{\bf Claim}
\theoremstyle{definition}
\newtheorem{example}{\bf Example}[section]
\theoremstyle{remark}
\newtheorem{remark}{Remark}[theorem]
\newcommand{\comb}[2] {\mbox{$\left( { #1 \atop #2 } \right)$}}
\newcommand{\card}[1]{\mbox{$\mid #1 \mid$}}
\newcommand{\ovx}{\overline x}
\newcommand{\ovv}{\overline v}
\newcommand{\ovw}{\overline w}
\newcommand{\ovu}{\overline u}
\newcommand{\ovf}{\overline f}
\newcommand{\pf}{\noindent{\em Proof. }}

\renewcommand{\thefootnote}{\fnsymbol{footnote}}
\makeatletter
\renewcommand\section{\@startsection{section}{1}{\z@}%
                                  {-3.5ex \@plus -1ex \@minus -.2ex}%
                                  {2.3ex \@plus.2ex}%
                                  {\normalfont\large\scshape\centering}}

\renewcommand\subsection{\@startsection{section}{1}{\z@}%
                                  {-3.5ex \@plus -1ex \@minus -.2ex}%
                                  {2.3ex \@plus.2ex}%
                                  {\normalfont\scshape}}

\renewcommand\subsubsection{\@startsection{section}{1}{\z@}%
                                  {-3.5ex \@plus -1ex \@minus -.2ex}%
                                  {2.3ex \@plus.2ex}%
                                  {\normalfont\itshape}}

\makeatother
\begin{abstract}
	Let $n, r, k$ be positive integers such that $3\leq k < n$ and $2\leq r \leq k-1$. Let $m(n, r, k)$ denote the maximum number of edges an $r$-uniform hypergraph on $n$ vertices can have under the condition that any collection of $i$ edges, span at least $i$ vertices for all  $1 \leq i \leq k$. We are interested in the asymptotic nature of $m(n, r, k)$ for fixed $r$ and $k$ as $n \rightarrow \infty$. This problem is related to the forbidden hypergraph problem introduced by Brown, Erd\H{o}s, and S\'{o}s and very recently discussed in the context of combinatorial batch codes. In this short paper we obtain the following results.
\begin{enumerate}[(i)]
\item  Using a result due to Erd\H{o}s we are able to show $m(n, k, r) = o(n^r)$ for $7\leq k$, and $3 \leq r \leq k-1-\lceil\log k \rceil$. This result is best possible with respect to the upper bound on $r$ as we subsequently show through explicit construction that for $6 \leq k$, and $k-\lceil \log k \rceil \leq r \leq k-1, m(n, r, k) = \Theta(n^r)$.\par
This explicit construction improves on the non-constructive general lower bound obtained by Brown, Erd\H{o}s, and S\'{o}s for the considered parameter values. 
\item For $2$-uniform CBCs we obtain the following results.
\begin{enumerate}
\item We provide exact value of $m(n, 2, 5)$ for $n \geq 5$.
\item Using a result of Lazebnik {\em et al.} regarding maximum size of graphs with large girth, we improve the existing lower bound on $m(n, 2, k)$ ($\Omega(n^{\frac{k+1}{k-1}})$) for all $k \geq 8$ and infinitely many values of $n$.
\item We show $m(n, 2, k) = O(n^{1+\frac{1}{\lfloor\frac{k}{4}\rfloor}})$ by using a result due to Bondy and Simonovits, and also show $m(n, 2, k) = \Theta(n^{\frac{3}{2}})$ for $k = 6, 7, 8$ by using a result of K\"{o}vari, S\'{o}s, and Tur\'{a}n.
\end{enumerate}
\end{enumerate}

\end{abstract}
\section{Preliminaries}
\subsection*{Hypergraphs and  Tur\'an numbers} We briefly make some definitions, and set up our notation. 

A hypergraph $F$ is a tuple $F := (V, \mathcal{F})$, where $V$ is a set of {\em vertices} and $\mathcal{F}$ is a family of subsets of $V$. Sets of $\mathcal{F}$ are called {\em edges} of the hypergraph and cardinality of $\mathcal{F}$ is called {\em size} of the hypergraph. A hypergraph $\mathcal{F}$ is called $r$-{\em uniform} if each of its edges has cardinality $r$. For a vertex $x \in \mathcal{V}$, we will denote by $deg(x)$, the number of edges in $\mathcal{F}$ containing $x$. Further, by $K_n^{(r)}$ we will denote the complete $r$-uniform hypergraph on $n$ vertices, and by $K^{(r)}(l, \ldots, l)$ we will denote the complete $r$-uniform $r$-partite hypergraph with $l$ vertices in each part. We denote by $K(s, t)$ the complete bipartite graph with partition sets of size $s$ and $t$ respectively and by $C_i$, a cycle of length $i$.\par
Let $\mathcal{H}$ be a family of $r$-uniform hypergraphs. By the  {\em Tur\'an number} of the family $\mathcal{H}$ denoted by $ex(n, \mathcal{H})$,  we mean the maximum size of an $r$-uniform hypergraph on $n$ vertices that does not contain a copy of any of the hypergraphs of $\mathcal{H}$ as a subgraph. 
\subsection*{Combinatorial Batch Codes and an extremal problem}
The notion of {\em Batch Codes} was introduced in \cite{YuKuOsSa} as an abstraction of a load balancing problem in a distributed database setup. Loosely speaking, an $(m, N, k, n, t)$\footnote{In batch code literature, number of data items is denoted by $n$ and number of servers is denoted by $m$. In this article, we deviate from this and reverse the roles of $m$ and $n$ to make it consistent with the common notations used in the hypergraph setting}-batch code models the problem of storing $m$ data items into $n$ servers in such a way that any $k$ \footnote{In this paper, we will refer to this parameter as the retrievability parameter.}of the $n$ data items may be retrieved by reading at most $t$ \footnote{In this article, we will exclusively consider the case of $t=1$ as that seems to capture the essence of the problem and this is the case for most of the work done in this area. So, henceforth, we will not mention this parameter in any expression with the understanding that $t=1$ case is considered.} items from each server and the overall storage to be limited to $N$. {\em Combinatorial Batch Codes} (CBCs), also introduced in \cite{YuKuOsSa}, and subsequently studied in \cite{PaStWe}, \cite{BrKiMeSc}, \cite{BuTu1,BuTu2,BuTu3}, \cite{BhRuRo}, is a subclass which models the scenario when each of the $N$ stored data items is a copy of each of the $m$ input data items, i. e., the $m$ input data items are replicated among $n$ servers. This restriction makes the problem a purely combinatorial one which can be studied in the setting of a hypergraph. Without providing further details (cf. \cite{PaStWe, BuTu1, BhRuRo}) we state the following theorem of \cite{PaStWe} which characterizes CBCs in the setting of a hypergraph.
\begin{theorem}[\cite{PaStWe}]
	A hypergraph $(\mathcal{V}, \mathcal{F})$ represents an $(m, N, k, n)$-CBC if $\lvert \mathcal{V} \rvert = n$, $\lvert \mathcal{F} \rvert = m$, $\sum_{F \in \mathcal{F}} \lvert F \rvert = N$ and every collection of $i$ edges of $\mathcal{F}$ contains at least $i$ vertices for $1 \leq i \leq k$.
	\label{cbcth}
\end{theorem}
Henceforth, we will refer to the hypergraph representing a CBC  as a CBC with corresponding parameters. Naturally a CBC will be termed $r$-uniform if the corresponding hypergraph is $r$-uniform.\par

The problem that we will address in this article is that of maximizing the number of data items ($m$) of a uniform CBC for given values of the number of servers ($n$) and retrievability parameter ($k$). Equivalently, and more formally we have the following extremal problem.
\begin{quote}
	{\em Let $n, r, k$ be positive integers such that $3\leq k < n$ and $2\leq r \leq k-1$. Determine $m(n, r, k)$, the maximum number of edges a $r$-uniform hypergraph on $n$ vertices can have under the condition that any collection of $i$ edges, $1 \leq i \leq k$, span at least $i$ vertices.}
\end{quote}
This is a forbidden hypergraph problem, where we have the following family of forbidden hypergraphs.
\begin{align}
	\label{forbset1}
\mathcal{H}_r(k) = \{H : H\  \mbox{\  is an $r$-uniform hypergraph with $i$ edges and $<i$ vertices for $1\leq i \leq k$}\}
\end{align}
We are interested in the asymptotic nature of $ m(n, r, k) = ex(n, \mathcal{H}_r(k))$ for fixed $r$ and $k$ as $n \rightarrow \infty$.\\
 This type of extremal problem was introduced by Brown, Erd\H{o}s, and S\'{o}s in \cite{BrErSo}, where the authors considered as forbidden family the following family of hypergraphs.
\begin{align}
\label{formbrerso}
\mathcal{H}_r(p, q) = \{H: H \mbox{ is an $r$-uniform hypergraph with $p$ vertices and $q$ edges } \}
\end{align} 
 They showed, through non-constructive arguments, the following lower bound.\\
\begin{align}
\label{brersolower}
ex(n, \mathcal{H}_r(p, q)) = \Omega(n^{\frac{rq-p}{q-1}})
\end{align}
We make the following observations to simplify the problem that we are considering in this article, i. e., finding the asymptotic nature of $m(n, r, k)$ for fixed $r$ and $k$ as $n \rightarrow \infty$.\\  
\begin{remark}
\label{rem1}
\begin{enumerate}
	\item The parameter $r$ considered in this article is  a constant independent of $n$, and since we are interested in the asymptotic nature of $m(n, r, k)$, we shall assume that all the hypergraphs studied in this article do not have multiple copies of any $r$-set, i. e., the considered hypergraphs are all simple. Indeed, observing that for the forbidden hypergraphs considered above, any edge can have at most $r$  copies, we are guaranteed to have at least a fraction of $\frac{1}{r}$ of the maximal possible size of an extremal hypergraph with repeated edges, and hence the asymptotic estimate for $m(n, r, k)$ does not change.
	\item If a collection of $\mathcal{F}_i$ of $i$ distinct edges span $<i$ vertices, then there is a subcollection $\mathcal{F}_j \subseteq \mathcal{F}_i$ of $j$ distinct edges that span exactly $j-1$ vertices.
	\item Though it does not find any significant role in the sequel, we mention here for the sake of preciseness that in a  simple hypergraph, a collection of $2$ distinct $r$-edges span at least $r+1$ vertices and a collection of $r+2$ distinct $r$-edges span at least $r+2$ vertices. 

\end{enumerate}
\end{remark}
From the above observations, the family of forbidden hypergraphs of (\ref{forbset1}) can be equivalently defined as below.
\begin{align}
	\label{forbset2}
\mathcal{H}_r(k) = \{H : H \mbox{\ is a simple $r$-uniform hypergraph with $i$ edges and $i-1$ vertices for $r+3\leq i \leq k$}\}
\end{align}

A lower bound for $m(n, r, k)$ was given in \cite{YuKuOsSa}, where the authors obtained the following result using a simple probabilistic argument:
\begin{align}
\label{lower1}
m(n, r, k) = \Omega(n^{r-1}).
\end{align}
In \cite{PaStWe}, the authors using a method (method of alteration, same as in the proof of (\ref{brersolower})) of \cite{BrErSo}, improved this lower bound: 
\begin{align}
\label{lower2}
m(n, r, k) = \Omega(n^{\frac{kr}{k-1}-1}).
\end{align}
On the other hand, in \cite{PaStWe}, the authors obtained the following upper bound:
\begin{align}
\label{upper}
m(n, r, k) \leq \frac{(k-1)}{\binom{k-1}{r}} \binom{n}{r}.
\end{align}
Now, it is trivial to see that this bound is met exactly for $r=1$. In \cite{PaStWe}, it was shown by explicit construction, that this bound is tight (in exact terms) for the cases $r=k-1$ and $r= k-2$. Indeed,  $k-1$ copies of $K_{n}^{(k-1)}$ and $k-2$ copies of  $K_{n}^{(k-2)}$, respectively do the job. \par

So, (\ref{upper}) essentially shows an upper bound $O(n^r)$ for $m(n, k, r)$. We are interested in the values of $r$ in the range $2 \leq r \leq k-3$, for $k\ge 5$.\par
\vspace{0.2cm}

In this article we obtain the following results.
\begin{enumerate}[(i)]
\item In Section \ref{runiform}, we improve the upper bound (\ref{upper}) in an asymptotic sense. In particular, using a result due to Erd\H{o}s (\cite{Er1}) we are able to show $m(n, k, r) = o(n^r)$ for $7\leq k$, and $3 \leq r \leq k-1-\lceil\log k \rceil$. This result is best possible with respect to the upper bound on $r$ as we subsequently show through explicit construction that for $6 \leq k$, and $k-\lceil \log k \rceil \leq r \leq k-1, m(n, r, k) = \Theta(n^r)$.\par
This explicit construction improves on the general lower bound (\ref{brersolower}) obtained by Brown, Erd\H{o}s, and S\'{o}s for the parameters $p = k-1, q = k, k-\lceil \log k \rceil \leq r \leq k-1$, where $k \geq 6$. In this case their lower bound is the same as (\ref{lower2}), i. e., $\Omega(n^{\frac{kr}{k-1}-1})$. 
\item In Section 5, we deal with the graph case, i. e., $2$-uniform CBCs and obtain the following results.\par
\begin{enumerate}
\item We provide exact value of $m(n, 2, 5)$ for $n \geq 5$. 
\item Using a result of Lazebnik {\em et al.} \cite{LaUsWol} regarding maximum size of graphs with large girth we improve the existing lower bound (\ref{lower2}) ($\Omega(n^{\frac{k+1}{k-1}})$) of \cite{PaStWe} for all $k \geq 8$ and infinitely many values of $n$.
\item We show $m(n, 2, k) = O(n^{1+\frac{1}{\lfloor\frac{k}{4}\rfloor}})$ by using a result due to Bondy and Simonovits, and also show $m(n, 2, k) = \Theta(n^{\frac{3}{2}})$ for $k = 6, 7, 8$ by using a result of K\"{o}vari, S\'{o}s, and Tur\'{a}n.
 
\end{enumerate}
\end{enumerate}
\section{$r$-uniform case for $r \geq 3$}
\label{runiform}
We begin this section by stating the following result due to Erd\H{o}s ( which is a generalization of a result of K\"{o}vari {\em et al.} \cite{KoSoTu}) that will be crucial in our proof of Theorem \ref{thm1}.
\begin{theorem}[\cite{Er1}]
	Let $n> n_{0}(r, l), l>1$. Then for sufficiently large $C$ ($C$ is independent of $n, r, l$), 
	$$
	n^{r-\frac{C}{l^{r-1}}} < ex(n, K^{(r)}(l, \ldots, l)) \leq n^{r-\frac{1}{l^{r-1}}}.
	$$
	\label{erth1}
\end{theorem}
We first show that $m(n, k, r) = o(n^r)$ for $7 \leq k$, and $3 \leq r \leq k-1-\lceil \log k \rceil$. All the logarithms mentioned in this paper are to the base $2$.

\begin{theorem}
	Let $ k \geq 7 $, and $3 \leq r \leq k-1-\lceil \log k \rceil$. Then for sufficiently large $n$ ($n > n_0(r)$), $m(n, r, k) \leq n^{r- \frac{1}{2^{r-1}}}$.
	\label{thm1}
\end{theorem}
\pf Let $7 \leq u \leq k$, $1\leq  v \leq u - 2\lceil\log u\rceil$ be such that $r = u-v-\lceil \log u \rceil$. It is clearly possible to find such $u$, $v$ for the range of values of $r$ as in the hypothesis. Consider the $r$-uniform, complete $r$-partite hypergraph $\mathcal{H} = (\mathcal{V}, \mathcal{F})$, where
$$
\mathcal{V} := \{x_1,\ldots,x_{u-v-2\lceil\log u\rceil}, \ldots, x_{r}, y_1, \ldots, y_{u-v-2\lceil\log u \rceil}, \ldots, y_{r}\}
$$
and 
$$
\mathcal{F} := \bigg\{\{z_1, \ldots, z_r\}: z_i \in \{x_i, y_i\}, 1 \leq i \leq r\bigg\}.$$

Next, consider the $r$-uniform sub-hypergraph $\mathcal{H'} = (\mathcal{V'}, \mathcal{F'})$ of $\mathcal{H}$, where
$$
\mathcal{V} \supseteq \mathcal{V'} := \{x_1,\ldots,x_{u-v-2\lceil\log u\rceil}, \ldots, x_r,  y_{u-v-2\lceil\log u \rceil+1}, \ldots, y_r\}
$$
and 
$$
\mathcal{F'} := \bigg\{  \{x_1,\ldots,x_{u-v-2\lceil\log u\rceil}, z_{u-v-2\lceil\log u\rceil+1}, \ldots, z_r\} 
 : z_j \in \{x_j, y_j\}, u-v-2\lceil\log u\rceil+1 \leq j \leq r\bigg\}.
$$
Since $v \geq 1$,
\begin{align}
	\label{vertineq}
	\lvert \mathcal{V'} \rvert = u - v \leq u-1,  
\end{align}
and
\begin{align}
	\label{edgeineq}
\lvert \mathcal{F'} \rvert = 2^{\lceil \log u \rceil} \geq u. 
\end{align}
So, $\mathcal{H}$ is a forbidden hypergraph required in Theorem \ref{erth1} containing a forbidden hypergraph $\mathcal{H'}$ of the collection $\mathcal{H}_{r}(k)$ of (\ref{forbset1}). Hence, it follows from Theorem \ref{erth1} that for sufficiently large $n$, i. e., for $n > n_0(r)$, 
$m(n, r, k) \leq ex(n, \mathcal{H'}) \leq ex(n, \mathcal{H}) \leq n^{r- \frac{1}{2^{r-1}}}$.\qed
\vspace{0.2cm}

\par
A few remarks regarding the theorem are in order.\\

\begin{remark}
	\begin{enumerate}
		\item Each edge of $\mathcal{F'}$ has the fixed set of vertices $\{x_1,\ldots,x_{u-v-2\lceil\log u\rceil} \}$. This choice is arbitrary and any fixed set of $u-v-2\lceil\log u\rceil$ vertices $\{z_1, \ldots, z_{u-v-2\lceil\log u\rceil}\}$ could have been chosen maintaining the condition $z_j \in \{x_j, y_j\}, 1\leq j \leq u-v-2\lceil\log u\rceil$.
		\item 
		One can also see that the same construction with partite sets of size $l$, in conjunction with Theorem \ref{erth1} gives us a corresponding result for $r\leq k-1-(l-1)\lceil \log_l k \rceil$. 
		\item Inequalities (\ref{vertineq}) and (\ref{edgeineq}) are tight when $u$ is a power of $2$ and $v = 1$.
			In particular, when $k$ is a power of $2$ and $r = k-1- \log k$, we have $\lvert \mathcal{V'} \rvert = k-1$ and $\lvert \mathcal{F'} \rvert = k$. So, $k$ edges of $\mathcal{H'}$ span exactly $k-1$ vertices.
		\end{enumerate}
\end{remark}
\begin{theorem}
$m(n, r, k) = \Theta(n^r)$ for $6 \leq k$, $k - \lceil \log k \rceil \leq r \leq k-1$.
\label{thm2}
\end{theorem}
\pf Here we show $m (n, r, k) = \Omega(n^r)$ for the stated ranges of values of $r$ and $k$. This, together with $m(n, r, k) = O(n^r)$ from (\ref{upper}), would imply $m(n, r, k) = \Theta(n^r)$. Also, first we prove the above for $r = k - \lceil \log k \rceil$, as this turns out to be the tight case and the same argument can easily be applied for the rest of the range of values of $r$. Note that the cases $r = k-1$ and $r = k-2$ have already been settled in \cite{PaStWe}.\\
{\em Construction:} Consider the complete $r$-uniform, $r$-partite hypergraph $\mathcal{H} = (\mathcal{V}, \mathcal{F})$, where $\mathcal{V} = \mathcal{V}_1 \cup \mathcal{V}_2 \cup \ldots \cup \mathcal{V}_r$, such that $ \mathcal{V}_i \cap \mathcal{V}_j = \emptyset $ for $i \neq j$, and  $\lvert \mathcal{V}_i \rvert = \lfloor \frac{n+i-1}{r} \rfloor$, for $1 \leq i \leq r$. Clearly $\lvert \mathcal{F} \rvert = \Omega (n^r)$.\par
\begin{claim}
$\mathcal{H}$ does not contain a sub-hypergraph $\mathcal{H'} = (\mathcal{V'}, \mathcal{F'})$ such that $\lvert \mathcal{V'} \rvert= i-1 $ and  $\lvert \mathcal{F'} \rvert \geq i$ for $r+3 \leq i \leq k$.
\end{claim}
\pf First, we observe that any sub-hypergraph $\mathcal{H'} = (\mathcal{V'}, \mathcal{F'})$ such that $\lvert \mathcal{V'} \rvert= i-1 $ and  $\lvert \mathcal{F'} \rvert \geq i$ for $r+3 \leq i < k$, there is another sub-hypergraph $\mathcal{H''} = (\mathcal{V''}, \mathcal{F''})$, such that $\mathcal{H'} \subseteq \mathcal{H''} \subseteq \mathcal{H}$, $\lvert \mathcal{V''} \rvert= k-1 $, and  $\lvert \mathcal{F''} \rvert \geq k$. To get $\mathcal{H''}$ from $\mathcal{H'}$ we simply add $k-i$ edges to $\mathcal{F'}$, where each edge contains exactly one unique vertex not contained in $\mathcal{V'}$, i. e., these $k-i$ unique vertices are distinct for these distinct $k-i$ added edges. This is always possible due to the structure of $\mathcal{H}$ provided there are $k-i$ distinct vertices in $\mathcal{V} \setminus \mathcal{V'}$. But this can be safely assumed because $n$ is large enough (due to the asymptotic nature of the problem), in fact, $n \geq 2k$ would suffice. Due to this, it is sufficient to establish that $\mathcal{H}$ does not contain a sub-hypergraph $\mathcal{H'} = (\mathcal{V'}, \mathcal{F'})$ such that $\lvert \mathcal{V'} \rvert= k-1 $ and  $\lvert \mathcal{F'} \rvert \geq k$.\par  

Next, let us consider any $\mathcal{V'} \subseteq \mathcal{V'}$ with $\lvert \mathcal{V'} \rvert= k-1$.  Then $\mathcal{V'} = \mathcal{V'}_1 \cup \mathcal{V'}_2 \cup \ldots \cup \mathcal{V'}_r$, where $ \mathcal{V'}_i \subseteq  \mathcal{V}_i$, for $1 \leq i \leq r$. Also $\lvert  \mathcal{V'}_i \rvert \geq 1$ (for otherwise $\mathcal{F'}$ would be empty, trivially prooving the claim), $\Sigma_{i=1}^{r} \lvert  \mathcal{V'}_i \rvert = \lvert  \mathcal{V'} \rvert = k-1$, and $ \lvert \mathcal{F'} \rvert = \Pi_{i=1}^{r} \lvert \mathcal{V'}_i\rvert$. Here, it is important to note that $\lvert \{ \mathcal{V'}_i : \lvert\mathcal{V'}_i \rvert=1, 1\leq i \leq r\}\rvert\geq 2r-k+1 \geq 1$ for $6 \leq k$, and $r = k - \lceil \log k \rceil$.\par
Now, observe that maximum of $\lvert  \mathcal{F'} \rvert$ is attained when there are exactly $(k-r-1)$ $\mathcal{V'}_i$s, $1 \leq i \leq k$ with $\lvert \mathcal{V'}_i \rvert = 2$, and for the remaining $2r-k+1$ $\mathcal{V'}_i$s,  $\lvert \mathcal{V'}_i \rvert = 1$. This can be seen by the following shifting argument.\par
Without loss of generality let us assume that $\lvert\mathcal{V'}_1\rvert =1$ and $\lvert\mathcal{V'}_2\rvert = l$, where $l \geq 3$. Now, it is easy to see that by shifting a vertex from $\mathcal{V'}_2$ to $\mathcal{V'}_1$,  $\lvert\mathcal{V'}_1\rvert \lvert \mathcal{V'}_2 \rvert$ becomes $2(l-1)$, which is $> l$ for $l \geq 3$.\par
So, following the above argument, maximum of $\lvert  \mathcal{F'} \rvert$ is attained when $\lvert\mathcal{V'}_i\rvert$ are as equal as possible for $1 \leq i \leq r$, and since as observed earlier  we always have $\lvert \{ \lvert\mathcal{V'}_i \rvert=1, 1\leq i \leq r\}\rvert \geq 1$, it follows that $\lvert\mathcal{V'}_i \rvert \in \{1, 2\}$ for $1 \leq i \leq r$. This implies  $\lvert  \mathcal{F'} \rvert \leq 2^{k-r-1} = 2^{\lceil \log k \rceil -1} \leq k-1$. This proves the claim.\par
Now, applying the same argument for the cases $k - \lceil \log k \rceil < r \leq k-1$, one can easily see that $\lvert  \mathcal{F'} \rvert \leq 2^{k-r-1} < 2^{\lceil \log k \rceil -1} \leq k-1$. Hence, the theorem. \qed 

\section{$2$-uniform case}
\label{2uniform}
For $2-$uniform (graph) CBCs, we provide the following improvements over existing results.

\vspace{0.2cm}
\subsubsection{An Exact Result}
\begin{theorem}
	$m(n, 2, 5) = \lfloor\frac{n^2}{4} \rfloor$ for $n \geq 5$.
\end{theorem}
\pf We first remark that in \cite{PaStWe}, Paterson {\em et al.}  observed that the complete bipartite graph on $n$ vertices satisfies the batch condition for $k = 5$, and hence concluded that $m(n, 2, 5) \geq \lfloor\frac{n^2}{4} \rfloor$. Hence it suffices to prove that $m(n,2,5)\le\lfloor\frac{n^2}{4} \rfloor$. In particular we only need to show that any graph with $n$ vertices and at least $\lfloor\frac{n^2}{4} \rfloor + 1$ edges contains a subgraph with $4$ vertices and $5$ edges.\par	
We prove this by induction on $n$.  This is clearly true for $n=4$. Now suppose we have a graph with $n$ vertices satisfying the given condition. We may assume that the graph has exactly $\lfloor\frac{n^2}{4} \rfloor + 1$ edges by throwing any extra edges away. Observe that the graph contains a vertex of degree at most $\lfloor \frac{n}{2} \rfloor$. Removing this vertex along with all its incident edges leaves a graph with $n-1$ vertices and at least $\lfloor \frac{(n-1)^2}{4}\rfloor+1$ edges, which by the induction hypothesis contains a subgraph with $4$ vertices and $5$ edges. \qed
\par
\vspace{0.2cm}
\subsubsection{Improvement of the lower bound}
For improvement of the lower bound we need the following lemma whose proof, although not difficult, is provided here for the sake of completeness.
\begin{lemma}
	\label{grbnd}
	Let $k\ge 6$ be a positive integer. If a graph has $k$ edges  and at most $k-1$ vertices then it has girth at most $\lfloor\frac{2k}{3}\rfloor$. This bound is tight.
\end{lemma}
\pf For $k=6$ the statement is clear. Suppose the statement does not hold for some $k>6$. Pick $k$ minimum so that the statement does not hold, i.e.,  suppose that for some $k$ we have a graph $G$ with $k$ edges, and at most $k-1$ vertices, and the firth of $G$ is at least $\geq \lfloor \frac{2k}{3}\rfloor+1$. We may assume that the graph is connected. Since $G$ has $k$ edges and at most $k-1$ vertices, it contains at least $2$ distinct cycles, $C_1,C_2$, say.  Now, if $C_1$ and $C_2$ are edge disjoint then one of them will have length at most $\lfloor \frac{k}{2} \rfloor < \lfloor \frac{2k}{3}\rfloor$ which contradicts the assumption regarding $k$.\par
If $C_1$ and $C_2$ are not edge  disjoint let $l_0=|E(C_1)\cap E(C_2)|$ be the number of common edges and $l_1, l_2$ be the number of edges that exclusively belonging to the $C_1$, and $C_2$ respectively. Clearly, $l_0+l_1=|E(C_1)|$, and $l_0+l_2=|E(C_2)|$. Now,  consider the subgraph of $G$ consisting of the edges of $C_1\Delta C_2$;  every vertex in this subgraph has even degree, so in particular it contains a cycle.  By the assumption on the girth of $G$,
 we have
$$ l_i+l_j \geq \lfloor \frac{2k}{3}\rfloor+1,$$ for  $0\leq i < j\leq 2$, which gives 
$$ l_0+l_1+l_2 > k,$$ a contradiction. \par
This bound is best possible since we may consider two vertices joined by three pairwise vertex-disjoint paths of length $\frac{k}{3}$ (for $k$ a multiple of 3) to give a tight example.\qed
\par
\vspace{4pt}
Next we mention the following result of Lazebnik {\em et al.} and apply it to improve the lower bound \ref{lower2} on $m(n, 2, k)$ of \cite{PaStWe}, which in this case is $\Omega(n^{\frac{k+1}{k-1}})$, for all $k$ and infinitely many values of $n$.
\begin{theorem}[\cite{LaUsWol}]
        \label{lauswol}
        For $s\geq 2$, $ex(n, \{C_3 , C_4 , \ldots, C_{2s+1} \}) =\Omega(n^{1+ \frac{2}{3s-3+\epsilon}})$ for infinitely many values of $n$, where $\epsilon= 0$ if $s$ is odd, and $= 1$ if $s$ is even.
        \begin{corollary}
Let $k \geq 8$ then
$$m(n, 2, k) = \left\{
\begin{array}{l l}
\Omega(n^{\frac{k-3}{k-5}}) & \quad \text{if $k = 5 \mod 6$ }\\
\Omega (n^{\frac{k-2}{k-4}}) & \quad \text{if $k = 2 \mod 6$ or $k = 4 \mod 6$}\\
\Omega (n^{\frac{k-1}{k-3}}) & \quad \text{if $k = 1 \mod 6$ or $k = 3 \mod 6$}\\
\Omega (n^{\frac{k}{k-2}}) & \quad \text{if $k = 0 \mod 6$}
\end{array}
\right.$$
for infinitely many values of $n$.
        \end{corollary}

\end{theorem}
\pf The proof follows directly from Lemma \ref{grbnd} and Theorem \ref{lauswol}. We remark here that for the cases where $k = 0 \mod 3$ or $k = 1 \mod 3$ the bounds of the corollary may be improved as Lemma \ref{grbnd} requires the girth to be  $\lfloor\frac{2k}{3}\rfloor + 1$, which is odd in these cases. Whereas the bounds obtained were by applying Theorem \ref{lauswol} for graphs of girth  $\lfloor\frac{2k}{3}\rfloor + 2$.
\vspace{0.2cm}

\subsubsection{Improvement of the upper bound}

Next, we show improve the upper bound on $m(n, 2, k)$.
We begin with the following theorem due to Bondy and Simonovits.
\begin{theorem}(\cite{BoSi})
        If in a graph of order $n$, the number of edges $> 100kn^{1+\frac{1}{k}}$, then the graph contains a $C_{2l}$ for every $l \in [k, kn^{\frac{1}{k}}]$.
        \label{BoSi}
\end{theorem}
\begin{theorem}
       For $k\geq 4, m(n, 2, k) = O(n^{ 1+ \beta})$, where $\beta = \frac{1}{\lfloor \frac{k}{4}\rfloor}$.
\label{genupr}
\end{theorem}
\pf Here we show that $m(n, 2, k) \leq 100k n^ {1+ \beta}$. Consider a graph $G$ with $ 100k n^ {1+\beta}$ edges. Now, it is well known that any graph has a subgraph whose minimum degree is at least half of the average degree of the original graph. Hence, $G$ has a subgraph $H$ with minimum vertex degree at least $100k n^ {\beta}$. \par
Now, Theorem \ref{BoSi} implies that $H$ has a cycle $C$ of length at most $2 \lfloor \frac{k}{4} \rfloor$. Let $v \in C$ be an arbitrary vertex, and consider all the walks of length $\lfloor\frac{k}{4}\rfloor$ in $H$ starting at $v$ in which no edge repeats consecutively in the walk. It is easy to see that the number of such walks is
$$ 100kn^{\beta}(100kn^{\beta}-1)^{\lfloor \frac{k}{4}\rfloor -1} > n.$$  Consequently, there is a vertex $v'$, such that at least two distinct walks of length $\lfloor \frac{k}{4}\rfloor$ starting at $v$, terminate at $v'$. These two walks along with $C$ give rise to a forbidden configuration consisting of at most $k$ edges spanning at most $k-1$ vertices.\qed

Theorem \ref{genupr} leads to trivial upper bound $O(n^2)$ on $m(n, 2, k)$ for $k = 6, 7$. However, it turns out that we can improve this trivial upper bound on $m(n, 2, k)$for $k = 6, 7$ by the following well-known theorem due to K\"{o}vari {\em et al.} whose generalization is Theorem \ref{erth1}. 
\begin{theorem}[\cite{KoSoTu}, see also \cite{Bol}]
Suppose $2 \leq s$, $2 \leq t$. Then
$$ ex(n, K(s, t)) \leq \frac{1}{2} (s-1)^{\frac{1}{t}}(n-t+1) n^{1-\frac{1}{t}} + \frac{1}{2}(t-1)n.$$
\end{theorem}
\begin{corollary}
\label{kosotu}
  $m(n, 2, k) = \Theta(n^{\frac{3}{2}})$ for $k = 6, 7, 8$.
\end{corollary}
\pf \begin{enumerate}[(a)]
\item Theorem \ref{kosotu} clearly implies an upper bound of $O(n^{\frac{3}{2}})$ for $ex(n, K(s,2))$. More precisely, it implies there is a constant $c_{s,2}$ such that for all $n > n_0$, any graph of order $n$ with more than $c_{s, 2}n^{\frac{3}{2}}$ edges will contain a $K(s, 2)$. Now, $s = \lceil\frac{k}{2}\rceil, (k \geq 6)$ serves our purpose. Because $K(\lceil\frac{k}{2} \rceil, 2)$ has $\geq k$ edges and $\leq k-1$ vertices for $k \geq 6$, and hence contains the forbidden structure of $k$ edges spanning $\leq k-1$ vertices as subgraph. This implies $m(n, 2, k) = O(n^{\frac{3}{2}})$ for $k \geq 6$.\par
	To prove the asymptotic tightness of the upper bound for the cases $k =6, 7, 8$,  we note that Lemma \ref{grbnd} implies a graph which is $\{C_3, C_4, C_5\}$-free, gives rise to $2$-uniform CBC with retrievability parameter $k$  for $k \leq 8$. Now, it is well-known that for $q$ a prime power, the incidence graph of $PG(2,q)$ is a $q+1$-regular bipartite graph  with $2(q^2+q+1)$ vertices and girth at least $6$. In fact, and it was shown in \cite{ErReSo}, for sufficiently large $n$ this construction leads to a graph on $n$ vertices having $\Omega(n^{\frac{3}{2}})$ whose girth is at least $6$. Incidentally, this construction also improves on the non-constructive lower bound of (\ref{lower2}) for $m(n, 2, 7)$ and $m(n, 2, 8)$. So, finally we have $m(n, 2, k) = \Omega(n^{\frac{3}{2}})$ for $k = 6, 7, 8$.\par

\end{enumerate}
 \qed
\par
\section{Concluding Remarks}
While the draft was under preparation, the authors were informed by C. Bujt\'{a}s and Z. Tuza \cite{BuTu5} about their simultaneous and independent work in similar direction. The family of forbidden configurations they consider is the following.  

$$\mathcal{H}^{r}(k, q) = \{H: H \mbox{ is $r$-uniform } \wedge \lvert E(H)\rvert - \lvert V(H) \rvert = q+1 \wedge 1 \leq \lvert E(H) \rvert \leq k\},$$
 where $r \geq 2, q \geq -r+1, k \geq q+r+1$ are fixed integers, and $V(H)$ and $E(H)$ denote respectively the number of vertices and edges of the hypergraph $H$.\par
Their results improve on the upper bound on $m(n, r, k)$ given in Theorem \ref{thm1} for some ranges of values of $r$. 
Further, their results show that $m(n, 2, k) = O(n^{1+\frac{1}{\lfloor \frac{k}{3}\rfloor}})$ which clearly improves on our upper bound of $ O(n^{1+\frac{1}{\lfloor \frac{k}{4}\rfloor}})$, and also imply, together with Corollary \ref{lauswol}, the following asymptotically exact results:
$$m(n, 2, k) = \Theta(n^{\frac{4}{3}}), \textrm{\ for\ } k = 9, 10 ,11.$$

\end{document}